\documentclass{elsart3}

 \usepackage{epsfig}

\usepackage{amssymb}

\begin{document}

\begin{frontmatter}



\vspace*{-2.0cm}

\title{The BAIKAL Neutrino Experiment: From NT200 to NT200+}


\author[a]{V. Aynutdinov},
\author[a]{V. Balkanov},
\author[d]{I. Belolaptikov},
\author[a]{L. Bezrukov},
\author[a]{D. Borschov},
\author[b]{N. Budnev},
\author[a]{I. Danilchenko},
\author[a]{Ya. Davidov},
\author[a]{G. Domogatsky},
\author[a]{A. Doroshenko},
\author[b]{A. Dyachok},
\author[a]{Zh.-A. Dzhilkibaev},
\author[f]{S. Fialkovsky},
\author[a]{O. Gaponenko},
\author[d]{K. Golubkov},
\author[b]{O. Gress},
\author[b]{T. Gress},
\author[b]{O. Grishin},
\author[a]{A. Klabukov},
\author[h]{A. Klimov},
\author[a]{S. Klimushin},
\author[b]{A. Kochanov},
\author[d]{K. Konischev},
\author[a]{A. Koshechkin},
\author[c]{L. Kuzmichev},
\author[f]{V. Kulepov},
\author[a]{B. Lubsandorzhiev},
\author[a]{S. Mikheyev},
\author[e]{T. Mikolajski},
\author[f]{M. Milenin},
\author[b]{R. Mirgazov},
\author[c]{E. Osipova},
\author[b]{A. Pavlov},
\author[b]{G. Pan'kov},
\author[b]{L. Pan'kov},
\author[a]{A. Panfilov},
\author[a]{D. Petukhov},
\author[d]{E. Pliskovsky},
\author[a]{P. Pokhil},
\author[a]{V. Poleschuk},
\author[c]{E. Popova},
\author[c]{V. Prosin},
\author[g]{M. Rozanov},
\author[b]{V. Rubtzov},
\author[b]{Yu. Semeney},
\author[a]{B. Shaibonov},
\author[c]{A. Shirokov},
\author[e]{Ch. Spiering},
\author[b]{B. Tarashansky},
\author[d]{R. Vasiliev},
\author[e]{R. Wischnewski\corauthref{cor}},
\corauth[cor]{Corresponding author.}
\ead{ralf.wischnewski@desy.de}
\author[c]{I. Yashin},
\author[a]{V. Zhukov}

\address[a]{Institute for Nuclear Research, 60th October Anniversary pr. 7a, 
Moscow 117312, Russia}
\address[b]{Irkutsk State University, Irkutsk, Russia}
\address[c]{Skobeltsyn Institute of Nuclear Physics  MSU, Moscow, Russia}
\address[d]{Joint Institute for Nuclear Research, Dubna, Russia}
\address[e]{DESY, Zeuthen, Germany}
\address[f]{Nizhni Novgorod State Technical University, Nizhni Novgorod, 
Russia}
\address[g]{St.Petersburg State Marine University, St.Petersburg, Russia}
\address[h]{Kurchatov Institute, Moscow, Russia}

\begin{abstract}

The Baikal Neutrino Telescope has been operating 
in its NT200 configuration since April, 1998.
The telescope has been upgraded in April, 2005 
to the 10\,Mton scale detector NT200+.
It's main physics goal is the detection of signals from 
high energy neutrino cascades.
NT200+ reaches a 3-year sensitivity 
of $2 \times 10^{-7}$cm$^{-2}$s$^{-1}$sr$^{-1}$GeV
for an all-flavor diffuse cosmic 
$E^{-2}$  
neutrino flux 
for energies  
10$^2$~TeV~$\div$~10$^5$~TeV.

Desgin and sensitivity of NT200+ are described.
NT200+ is forming the basic 
building block
of a future km3-scale (Gigaton-Volume) Baikal Telescope.
Research and development work on that next stage detector has 
started.

\end{abstract}


\end{frontmatter}

\section{Introduction}    


\begin{figure}[t]
\begin{center}
\epsfig{file=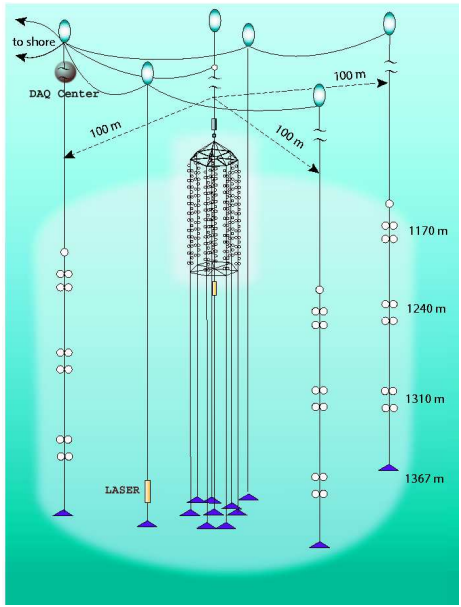,width=0.8\linewidth}
\caption{
The upgraded Baikal Telescope NT200+ : 
The old NT200 surrounded
by three external long strings at 100\,m radius from the center.
Also indicated: external laser and DAQ center.
\label{FIG_NT200}}
\end{center}
\end{figure}

The Baikal Neutrino Telescope  is operated in Lake 
Baikal, Siberia,  at a depth of {1.1~km}. 
Deep Baikal water is characterized by an absorption length of $L_{abs}(480 $nm$) =20\div
 24$ m,
a scattering length of $L_s =30\div 70$ m and a strongly anisotropic scattering function
with a mean cosine of scattering angle $0.85\div 0.9$ \cite{APP1}. 

The first stage telescope configuration  NT200
\cite{APP1} was put into 
permanent operation 
on April 6th, 1998 and consists of 192 optical modules (OMs).
An umbrella-like frame carries  8 strings,
each with 24 pairwise arranged OMs 
(see central part of Fig. \ref{FIG_NT200}).
Four underwater electrical cables connect the
detector with the shore station. 
Each optical module
contains a 37-cm diameter {QUASAR}
- photomultiplier (PM) which has been developed 
specially for this project
\cite{OM2}. 
The two PMs of a pair are switched in local coincidence in order 
to suppress background from bioluminescence and PM noise; 
each pair defines a {\it channel}. 

The 
upgraded
telescope 
NT200+ was put into operation on April 9th, 2005. 
This configuration consists of the old NT200
telescope,
surrounded by three new, external strings
placed 100\,m away from NT200 (see Fig.\ref{FIG_NT200}).
With these new strings, the 
sensitivity of the Baikal telescope for very high energy neutrinos
increases by a factor 4.

With the NT200 telescope, a number 
of relevant physics results has been obtained so far:
searches for WIMPs, high energy atmospheric 
muon neutrinos and muons, 
relativistic and slow magnetic monopoles and diffuse 
extraterrestrial high energy neutrinos.
Reviews were given at this conference \cite{CAT_DJILKIB},
and in refs.\,\cite{NEU04,NANP05,ECRS04,APP2}. 
The NT200 
all-flavor
limit for a steady  
diffuse 
neutrino flux with $E^{-2}$ shape,
$E^2 \Phi = 8.1 \times 10^{-7}$cm$^{-2}$s$^{-1}$sr$^{-1}$GeV (20\,TeV $<$ E $<$ 50\,PeV), 
is among the most sensitive limits published so far by neutrino telescopes \cite{BAI_APP05}.

\begin{figure}[b]
\begin{center}
\epsfig{file=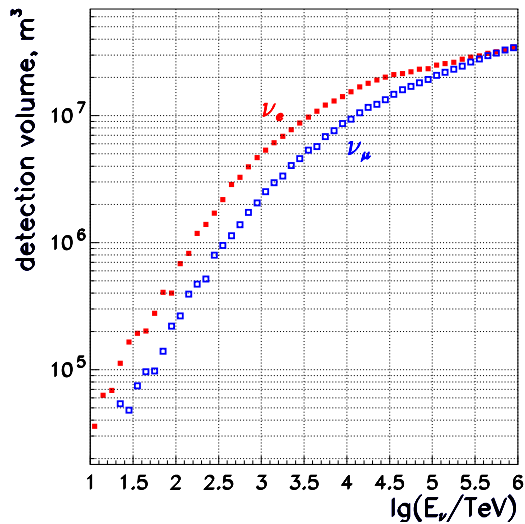,width=0.9\linewidth}
\caption{
Detection volume of NT200+
for $\nu_e$ and
$\nu_{\mu}$ events, after all cuts.
\label{FIG_EFFVOL}}
\end{center}
\end{figure}

For NT200, the detection strategy for high energy neutrino events
is based on 
a search for Cherenkov light 
from pointlike cascades in a Mton-scale sensitive volume below the 
detector, which is 
much exceeding the instrumented geometric volume (V$_{\rm{geo}}$(NT200)$\sim$ 0.1\,Mton).

With the upgrade to NT200+, the effective volume is now ``fenced'' by 
3 distant, long outer strings with only 36 OMs in total, 
which will 
give
physics information on high energy showers well exceeding that of NT200:
The long baseline allows to localize the shower position, and hence the
shower energy. This significantly 
improves the rejection capability against the physics background 
(mainly high energy atmospheric muons with em-showers, 
passing not far below NT200).
A measurement of the shower energy will be possible.
Fig.\ref{FIG_EFFVOL} gives the detection volume for NT200+ as function of 
neutrino energy for $\nu_e$ and $\nu_{\mu}$ events between 
10\,TeV and 1\,EeV.
The sensitivity to an 
$E^{-2}$-diffuse 
all-flavor 
neutrino flux  is
$2 \times 10^{-7}$cm$^{-2}$s$^{-1}$sr$^{-1}$GeV
(E $> 10^2$\,TeV, 3 years) \cite{BAI_APP05}.

In this paper, we describe the new NT200+ telescope, 
being the natural extension of NT200, and
it's 
data acquisition, control and calibration systems.
NT200+ 
will be used as the basic cell of 
a future km3-scale detector (Gigaton Volume Detector)
in Lake Baikal, which is briefly sketched.

\section {The NT200+ Telescope}

The upgraded telescope NT200+ was 
commissioned in April, 2005.
This new configuration consists of a
central part (the old telescope NT200) 
and three new, external strings (NT+), see Fig. \ref{FIG_NT200}. 
The external strings are 200~m long (140~m instrumented) 
and are placed at 100~meter
distance from the  center of NT200. Each string contains 12 OMs,
grouped in channels (OM pairs) like in NT200. 
The upper channels are at approximately the same depth as the bottom OMs of NT200,
adjacent channel distances are 20,50, 20, 30 and 20\,m from top to bottom
(for depths of upper, 3rd and lower channels see Fig.\ref{FIG_NT200}).
All channels are 
downlooking, 
except the lower two on each string (uplooking).

                                                                                
\begin{figure}[h]
\begin{center}
\epsfig{file=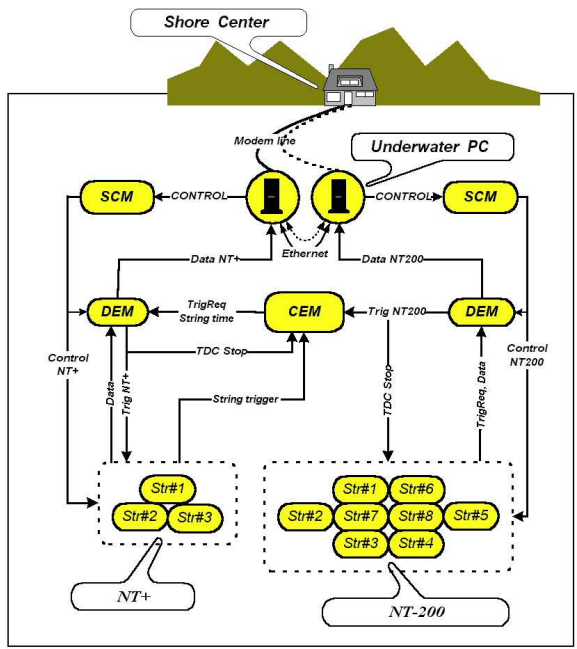,width=\linewidth}
\caption{
 Sketch of  data collection and slow control in NT200+:
 8-string telescope NT200 and 3 new outer strings (NT+),
 controlled from two underwater PCs.
\label{FIG_DAQ1}}
\end{center}
\end{figure}

\begin{figure}[h]
\vspace*{10mm}
\begin{center}
\epsfig{file=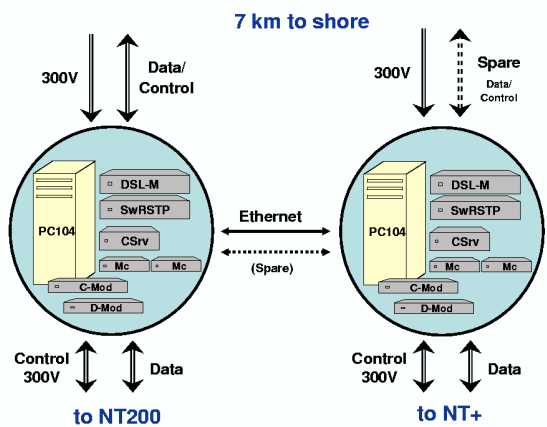,width=\linewidth}
\caption{
Sketch of the new NT200+ central DAQ/Control spheres 
with embedded PC104.
\label{FIG_DAQ2}}
\end{center}
\end{figure}

Two basic tasks had to be solved, 
in order to  
integrate 
NT200 and the external string subsystems  
into 
a united NT200+ detector: 
creation of a data acquisition and 
control system for the external strings, 
and providing 
a time synchronisation between the two subsystems
and a tagging of common events
(see also \cite{BAI_DAQ05}). 
Since a simple doubling of the data acquisition and control
system for NT200 and NT+ was compatible neither with the number 
of available cable connections to shore, nor  with future upgrades, 
we decided to significantly modernize the system by introducing 
for the first time embedded PCs with reliable industrial  Ethernet 
infrastructure underwater. 
For NT200+, all data and control cable connections of NT200 and the outer strings 
go 
to 
a new central control and readout unit ({\it DAQ center}) about 20\,m below 
surface (see Fig. \ref{FIG_NT200}),
where they are 
multiplexed to a single line to shore.  

All information necessary  
to combine 
NT200 and NT+ sub-events,
is provided by  
the 
common electronics module ({\it CEM}) specially designed 
for NT200+. {\it CEM} is located at 
the {\it DAQ center} 
and connected through 
coaxial cables with 
the NT200 and NT+ DAQ
subsystems. This module contains 
TDC units that measure time differences between  NT200 and NT+ triggers 
with a 
resolution of about 2\,ns. 
Also, all NT200 trigger pulses are 
counted and added to the NT+ data stream.
Figure \ref{FIG_DAQ1} sketches the DAQ and control system of NT200+, 
composed of the two  subsystems for NT200 and NT+.  String electronics 
modules ({\it SEM}) form the lower  level of NT200+ DAQ system. 
NT200 and NT+ strings ({\it Str}) contain two and one SEM, 
respectively. 
These units  translate trigger request signals from string 
channels to 
the
detector electronics 
module
({\it DEM}) or to {\it CEM} for NT+, and   
provide time and amplitude measurements for 
all
triggered channels.
A NT200 trigger is formed when
the number of fired channels $N_{hit}$ is at least $N_{min}$
within 500\,ns at {\it DEM}. 
$N_{min}$ is typically set to 3 or 4. 
The trigger signal is used as a common
stop for the TDCs of NT200 channels. 
For common operation with the 
external strings, the signal of the  NT200 trigger is sent through 1.2\,km
coaxial cable to {\it CEM}. 
The 
number of NT200 triggers 
within 
an experimental run  is 
recorded  
by a  counter in {\it CEM}. 
On each external NT+ string, 
triggers are formed as independent string-triggers, in case 
of at least 2 fired channels within 1000\,ns.  
String triggers are sent
to {\it CEM}, where the time difference between  string trigger and the 
trigger of NT200 is measured. 
This information is used to relate within an event 
the times of OMs in NT200 and the externals strings. 

NT200 and NT+ experimental  data are 
transfered 
to the shore center
through two  underwater 
PCs located in pressure glass spheres. 
Both underwater PC spheres
are nearly identical, their content is detailed in Fig. \ref{FIG_DAQ2}: 
a single board  PC/104 ({\it PC104:} Advantech-PCM9340),
a DSL-modem ({\it DSL-M:} FlexDSL-PAM-SAN, with hub and 
router), a managed Ethernet switch  ({\it SwRSTP:} RS2-4R, 
running RSTP protocols for the 
two-fold redundant  ethernet network between the PC spheres), 
an Ethernet-ComServer 
({\it CSrv:} WUT-58211, for PC-terminal emulation), 
two media-converters  ({\it Mc:} for coaxial 
connection to external control units) and the
experiment data and 
slow-control 
modems ({\it D-Mod}
and {\it C-Mod}). The connection to shore is by a single
DSL-Modem at a speed of up to 
2~Mbit/s. This full multiplexing of all data and control
streams 
through a single DSL-channel 
reduces 
the number of shore wires to two. Both PC spheres are
interconnected via two twisted pair ethernet
cables (main and hot spare). 
This underwater system works stable
since its first installation in 
2004. Using Linux throughout the system 
(for underwater PCs and shore station PCs)  
allows for  
easy remote maintenance and control from home institutions.

\section {NT200+ Laser Calibration}

Large volume underwater Cherenkov detectors need calibration of
the relative time-offsets between all light-sensors to a precision
of a few nsecs, since event reconstruction and classification
are based on the precise light arrival times.
For NT200+, calibration is done with a powerful external
laser light source
with up to $5 \times 10^{13}$ photons per pulse and 
nsec-pulse duration \cite{BAI_DAQ05},
which is 
located between two outer strings, see Fig.\ref{FIG_NT200}.
This ensures amplitudes of $\sim$100
photoelectrons on a few PMTs on each external string and on NT200.
High amplitudes minimize uncertainties due to light scattering.
\vspace*{.3cm} 

                                                                                
\begin{figure}[h]
\begin{center}
\epsfig{file=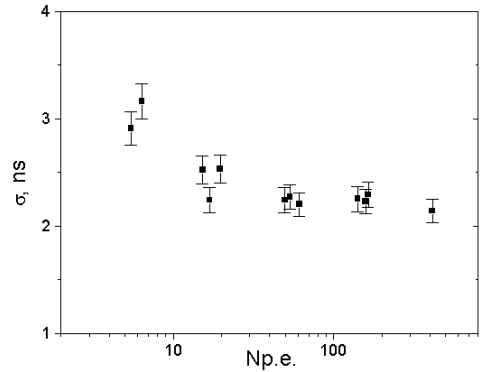,width=0.9\linewidth}
\caption{
 Measured NT200+ time resolution as function of PMT 
 amplitudes
 for laser calibration pulses.  
\label{FIG_LAS}}
\end{center}
\end{figure}

The NT200+ laser calibration unit is made of a powerful short-pulse
Nitrogen laser ($\lambda = 337$\,nm)
with about 100\,$\mu$J for $<$1\,nsec pulse duration, 
which is pumping a Coumarin dye laser at 480\,nm.
After passing through a computer-controlled attenuator disc,
the light is isotropized by a light diffuser ball, made of a
round-bottom flask filled with Silicone Gel (RTV-6156)
admixed with hollow micro-glass spheres at about 5\%-volume ratio
(S32 from 3M, with $\approx 40\mu$m diameter;
following an idea developed for the SNO detector \cite{SNO}).
The total loss of this isotropizing sphere 
is ${< 25\%}$.
All components are mounted into a 1\,m-long cylindrical glass 
pressure housing, which gives  
isotropic emission for more than the upper hemisphere.
The unit is installed at a depth of 1290\,m below surface and
operated in autonomous mode: after power-on from shore,
a series of pulses at various intensities is conducted. 

The final light output ranges from approximately 
$10^{12}$ to $5 \times 10^{13}$ photons/pulse, 
corresponding to shower energies from 10 PeV to 500 PeV.
The laser unit 
is
used, varying the total intensity, 
to calibrate pointlike shower vertex and 
energy reconstruction algorithms for energies up to 500\,PeV.

This laser unit
allows for
an independent performance check of 
the key elements of the NT200+ timing system. 
We performed the relative time synchronization
of all news strings and NT200, and  find the jitter of this 
to be less than 3\,nsec.  This jitter is due mainly 
to
the significant length (1.2~km) of synchronisation line  between
NT200 and external strings.  
The measured dependence of 
the 
relative time jitter on PMT amplitudes 
is presented in Fig. \ref{FIG_LAS} for
several pairs of channels of NT200 and external strings.

\section {A Gigaton Volume Detector at Baikal}

\begin{figure}[h]
\begin{center}
\epsfig{file=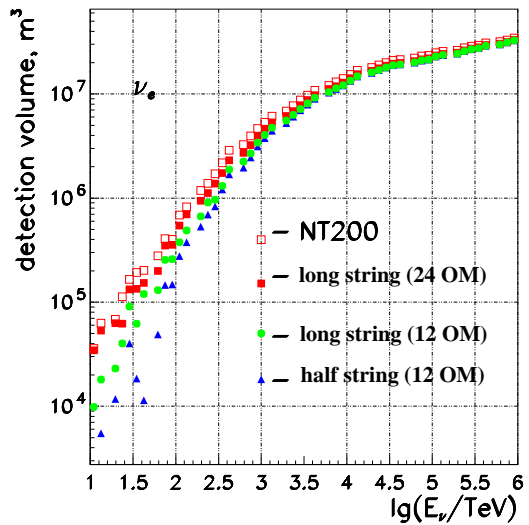,width=0.9\linewidth}
\caption{
Detection volumes of different configurations of the basic cell 
for a km3-size Baikal detector.
\label{FIG_GVD1}}
\end{center}
\end{figure}

MC simulations have shown that the detection volume of NT200$+$
for PeV cascades would vary only moderately, if NT200 as the
central part of NT200$+$ is replaced by a single string of OMs.
Figure \ref{FIG_GVD1} gives the detection volume for different configurations
as a function of cascade energy. 
The standard configuration of \mbox{NT200$+$} is marked 
by empty rectangles. 
The other configurations comprise a single string instead of NT200:
a standard string of 70\,m length and 24\,OMs (filled rectangles),
a  half string with 12\,OMs covering 35\,m (dots), and a 70\,m
long string sparsely equipped with 12\,OMs (triangles).
The configuration with the long  12-OM string shows  
an energy behavior very close to the one of NT200$+$. 

\begin{figure}
\begin{center}
\epsfig{file=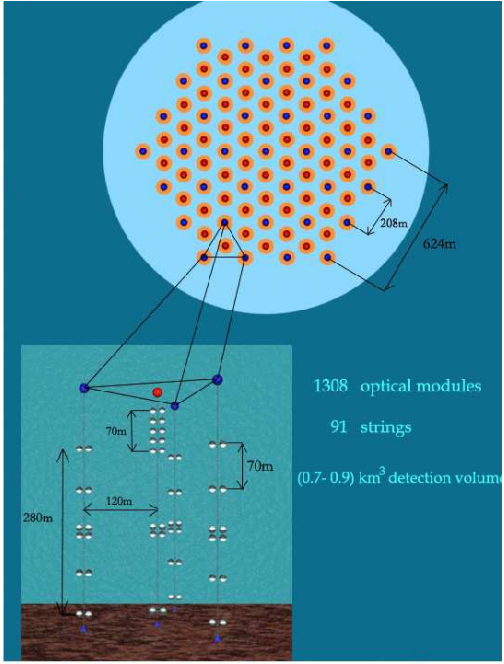,width=0.9\linewidth}
\vspace*{-0.6cm}
\caption{
Top view of the planned Baikal km3-detector (Gigaton Volume Detector).
Also shown is its basic cell: a ``minimized'' NT200+ telescope.  
\label{FIG_GVD2}}
\end{center}
\end{figure}

\begin{figure}[h]
\begin{center}
\epsfig{file=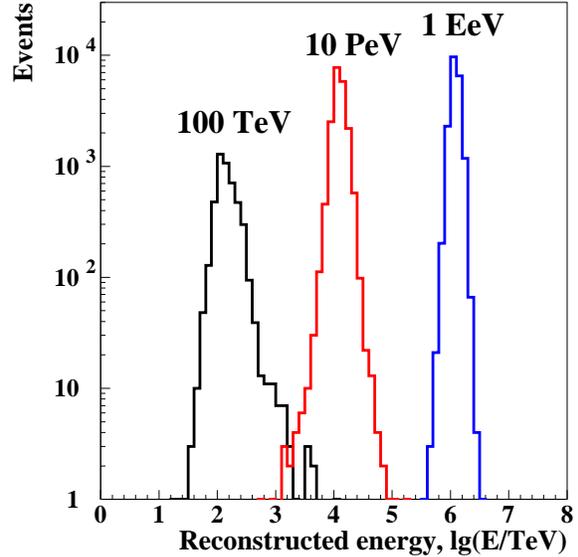,width=0.98\linewidth}
\caption{
Expected energy reconstrcution for cascades detected by the km3 detector.
\label{FIG_GVDEREC}}
\end{center}
\end{figure}

For neutrino energies above 100\,TeV,
such a configuration could be used as a basic 
building block
of a km3-scale or Gigaton Volume Detector (GVD).  
Rough estimations show that $0.7 \div 0.9$
Gton detection volume for neutrino induced high energy cascades 
may be achieved with about 1300 OMs arranged at 91 strings. 
A top view of GVD as well as 
a 
sketch of one basic subarray are shown in Fig.
\ref{FIG_GVD2}. 
The shower energy reconstruction capability 
is illustrated in Fig. \ref{FIG_GVDEREC}.
The 
physics program of
this
detector 
at very high energies covers the 
typical spectrum of cubic kilometer arrays.

                                                                                

\section {Conclusions and Outlook}

The deep underwater neutrino telescope NT200 in 
Lake Baikal 
has been
taking data since April 1998. 
A number of interesting results have been obtained,
based on the first years of NT200-operation.

The Baikal 
telescope has been significantly upgraded in 2005.  
The new 
telescope configuration NT200+ has a sensitivity 
better than 
$10^{-7}$cm$^{-2}$s$^{-1}$sr$^{-1}$GeV
for a diffuse 
$E^{-2}$  electron 
neutrino flux within the
energy range 10$^2$~TeV~$\div$~10$^5$~TeV.
NT200+ will search for neutrinos from AGNs, GRBs
and other extraterrestrial sources, neutrinos from cosmic ray
interactions in the Galaxy as well as high energy atmospheric muons
with E$_{\mu}>10$~TeV.

For the planned km3-detector in lake Baikal, R\&D-activities have recently been started.
Technical and physics experience with the new NT200+ detector 
will be an important part of this program.
With a Technical Design Report for the km3-detector 
scheduled for 2008, 
deployment will start in 2010.

\vspace{-0.5cm}
\ack
 This work was supported 
by NATO-Grant NIG-9811707(2005),
by the Russian Ministry of Education and Science, the 
German Ministry of Education and Research and 
the Russian Fund of Basic Research
(grants 05-02-17476, 04-02-17289, 04-02-16171, 05-02-31021, 05-02-16593), 
and by the Grant of the President of Russia NSh-1828.2003.2.


\bigskip\def\etal{{\em et al.}}

\end{document}